\documentclass[amsmath,amssymb,aps,twocolumn,showpacs,superscriptaddress,groupedaddress,nofootinbib,preprintnumbers]{revtex4}

\usepackage{url}
\usepackage{graphicx}
\usepackage{bm}
\usepackage{color}
\usepackage{enumitem}

\newcommand{\dd}{\mathrm{d}}

\newcommand{\ee}{{\rm e}}

\begin{document}

\author{Chul-Moon~Yoo}
\affiliation{
Division of Particle and Astrophysical Science,
Graduate School of Science, Nagoya University, 
Nagoya 464-8602, Japan
}

\author{Tomohiro~Harada}
\affiliation{
Department of Physics, Rikkyo University, Toshima,
Tokyo 171-8501, Japan
}

\author{Hirotada~Okawa}

\affiliation{
Waseda Institute for Advanced Study(WIAS), 
Waseda University, Shinjuku, Tokyo 169-8050, Japan
}

\title{Threshold of Primordial Black Hole Formation in Nonspherical Collapse}

\begin{abstract}
We perform (3+1)-dimensional 
simulations of primordial black hole (PBH) 
formation starting from the spheroidal
super-horizon perturbations. 
We investigate how the ellipticity (prolateness or oblateness)
affects the threshold of PBH formation in terms of the peak amplitude of 
 curvature perturbation.
We find that, in the case of the radiation-dominated universe, 
the effect of ellipticity on the threshold is negligibly 
small for the large amplitude of perturbations expected 
for PBH formation.
\end{abstract}

\preprint{RUP-20-12}

\maketitle

\section{Introduction}
\label{sec:intro}

The primordial black hole~(PBH) is a generic term used to refer to black holes 
which are generated in the early universe and are
not the final products of the
stellar evolution in late times. 
The possibility of PBH was firstly reported in Refs.~\cite{1967SvA....10..602Z,Hawking:1971ei}, 
and the remarkable characteristic is that, in contrast to black holes from stellar collapse, 
any mass of PBH is theoretically allowed in principle. 
The observational constraints are actively discussed and 
given in a broad mass range~(see e.g. Ref.~\cite{Carr:2020gox} for recent constraints). 
Despite the efforts to make constraints on the PBH abundance, 
PBHs are still viable and attractive candidates for a major part of dark matter~(e.g., see 
Ref.~\cite{Carr:2020gox} and references therein) 
or the origin of the binary black holes observed 
by gravitational waves~\cite{Abbott:2016blz,Sasaki:2016jop}. 
The most conventional scenario, which we suppose throughout this letter, 
is that PBHs are formed during the radiation-dominated universe 
as a result of gravitational collapse
of large amplitude of cosmological perturbations 
generated in the inflationary era.

When one estimates the PBH abundance, at least two ingredients are needed: one is the probability distribution for the 
parameters characterizing the initial inhomogeneity, and the other is the criterion for PBH formation. 
The criterion is often set for the amplitude of the initial inhomogeneity by using a threshold value
estimated through analytic~\cite{Carr:1975qj,Harada:2013epa} 
or numerical works~\cite{1978SvA....22..129N,1980SvA....24..147N,Shibata:1999zs,%
Niemeyer:1999ak,Musco:2004ak}
with spherical symmetry. 
Our aim in this letter is to estimate the effect of ellipticity on the 
threshold%
\footnote{
The growth of the anisotropic structure in the universe has been studied since a long time ago~(e.g., see Ref.~\cite{1981ApJ...250..432B}). 
A phenomenological approach to PBH formation can be seen in Ref.~\cite{Kuhnel:2016exn}. 
}. 

Recently, the spin of PBH 
has been 
attracting 
much attention~\cite{Chiba:2017rvs,Harada:2017fjm,DeLuca:2019buf,Mirbabayi:2019uph,Fernandez:2019kyb,He:2019cdb}. 
Once the typical value of the PBH spin is known, it can be compared with 
the observed spins of black holes 
such as black hole binaries observed by gravitational waves~\cite{Abbott:2016blz}. 
In order to clarify the spin distribution of PBHs, eventually, we need to perform 
full numerical simulations starting from relevant initial settings for PBH formation. 
In this letter, as a first step before discussing the spin, we perform the simulation of PBH formation 
with radiation fluid 
starting from a superhorizon-scale spheroidal inhomogeneity.

Throughout this letter, we use the geometrized units in which both
the speed of light and Newton's gravitational constant are unity, $c=G=1$.

\section{Initial data setting}
\label{sec:iniset}

In order to describe the initial superhorizon-scale inhomogeneity, 
we consider the situation $\epsilon:=k/(aH_{\rm b})\ll1$, 
where $1/k$ gives the characteristic comoving scale of the inhomogeneity, and $a$ and 
 $H_{\rm b}$ are the scale factor and Hubble expansion rate in the reference universe, respectively. 
Then 
the long-wavelength 
growing-mode solutions of all physical quantities up to the next leading order with respect to $\epsilon$ 
can be derived once the leading term of the curvature perturbation $\zeta$ is 
given as a function of the spatial coordinates $x^i$~\cite{Shibata:1999zs,Harada:2015yda}. 
The spatial metric is given by the reference flat metric multiplied by $\ee^{-2\zeta}a^2$ at the leading order. 
We use those long-wavelength solutions 
for the initial data in the numerical simulation. 

For concreteness, let us assume that the curvature perturbation $\zeta$ is a random Gaussian variable, 
and consider the probability distribution of the parameters characterizing the 
spatial profile of the curvature perturbation based on peak theory~\cite{1986ApJ...304...15B,Yoo:2018kvb}. 
First, we focus on a peak in $-\zeta$, and the
Taylor-series expansion up to the second order around this peak is
given as follows:
\begin{equation}
\zeta(X^i)=\zeta_0+\frac{1}{2}(\lambda_1 X^2+\lambda_2 Y^2+\lambda_3 Z^2),
\label{eq:Taylor}
\end{equation}
where $X^i=(X,Y,Z)$ are the appropriately rotated 
Cartesian coordinates. 
We can set $\lambda_1\geq\lambda_2\geq\lambda_3\geq0$ 
without loss of generality. 
Following Refs.~\cite{1986ApJ...304...15B,Yoo:2018kvb}, we introduce the following variables:
\begin{eqnarray}
\nu&=&-\zeta_0/\sigma_0,\\
\xi_1&=&(\lambda_1+\lambda_2+\lambda_3)/\sigma_2,\\
\xi_2&=&(\lambda_1-\lambda_3)/(2\sigma_2),\\
\xi_3&=&(\lambda_1-2\lambda_2+\lambda_3)/(2\sigma_2), 
\end{eqnarray}
where $\xi_2\geq\xi_3\geq-\xi_2$ and $\xi_2\geq0$ with $\sigma_n$ being
the $n$th-order 
gradient moment~\cite{1986ApJ...304...15B}. 
Throughout this letter, we assume $\sigma_n /k^n\ll1$. 
The probability density for these variables is given by~\cite{1986ApJ...304...15B,Yoo:2018kvb} 
\begin{equation}
P(\nu,\bm \xi)\dd \nu \dd \bm \xi=P_1(\nu,\xi_1)P_2(\xi_2,\xi_3)\dd \nu \dd \bm \xi, 
\label{eq:probden}
\end{equation}
where
\begin{eqnarray}
&&\hspace{-5mm}P_1(\nu,\xi_1)=\frac{1}{2\pi}\frac{1}{1-\gamma^2}
\exp\left[-\frac{1}{2}\left(\nu^2+\frac{(\xi_1-\gamma\nu)^2}{1-\gamma^2}\right)\right],~~~~~
\label{eq:p1}
\\
&&\hspace{-5mm}P_2( \xi_2,  \xi_3)=\frac{5^{5/2}3^2}{\sqrt{2\pi}} \xi_2
\left( \xi_2^2- \xi_3^2\right)
\exp\left[-\frac{5}{2}\left(3\xi_2^2+ \xi_3^2\right)\right]~~~~~
\end{eqnarray}
with $\gamma=\sigma_1^2/(\sigma_0\sigma_2)$. 
From this probability density, we find that 
there is no correlation between the two pairs $(\nu,  \xi_1)$ and $( \xi_2, \xi_3)$, 
and the typical values for $ \xi_2$ and $ \xi_3$, 
which characterize the ellipticity, are of the order of 1.

The dimensionless quantities which purely quantify 
the shape of the profile 
can be given by 
\begin{eqnarray}
\chi_1:=\xi_2/ \xi_1,~\chi_2:= \xi_3/ \xi_1. 
\end{eqnarray}
We note that, for the high-amplitude peaks which are relevant to PBH formation, 
according to Eq.~\eqref{eq:p1},  
typically we have 
\begin{equation}
 \xi_1 \sim \nu = -\zeta_0/\sigma_0 \gg 1, 
\label{eq:xi1typical}
\end{equation}
where we have assumed $\gamma\sim1$ and $|\zeta_0|\sim1$. 
Therefore the typical values of $\chi_1$ and $\chi_2$ for PBH formation are much smaller than 1, that is, 
the initial configuration of the system is 
typically highly spherically symmetric. 
Hence, from a cosmological point of view, 
our main concern is in PBH formation with small ellipticity. 

Because of the reflection symmetries of the profile \eqref{eq:Taylor} with respect to the surfaces $X^i=0$, 
we can restrict the numerical region to the cubic region $0\leq X^i\leq L$ ($i=1,2,3$)
as is adopted in Refs.~\cite{Yoo:2013yea,Yoo:2018pda}. 
Here we consider the following specific curvature perturbation profile characterized by 4 parameters 
$\mu$ and $k_i$:
\begin{equation}
\zeta=-\mu\exp\left[-\frac{1}{2}\left(k_1^2 X^2+k_2^2 Y^2+k_3^2Z^2\right)\right]W(R), 
\end{equation}
where $R=X^2+Y^2+Z^2$ and the function $W(R)$, 
which we do not specify here~(see Eq.~(24) in Ref.~\cite{Yoo:2018pda}), 
is introduced to smooth out the tail of the Gaussian profile on the boundary of 
the cubic region. 

In the simulation, we fix the square sum $\hat \xi_1$ of the wave numbers $k^i$ to $k^2$ as follows: 
\begin{equation}
\hat \xi_1:=\xi_1\sigma_2/\mu=k_1^2+k_2^2+k_3^2=k^2, 
\end{equation}
where we have used the relation $\mu=-\zeta_0$. 
Defining $\hat \xi_2:=\xi_2\sigma_2/\mu$ and $\hat \xi_3:=\xi_3\sigma_2/\mu$, we find 
$\chi_1=\hat \xi_2/k^2$ and $\chi_2=\hat \xi_3/k^2$, so that 
\begin{eqnarray}
3k_1^2&=&(\hat \xi_1+3\hat \xi_2+\hat \xi_3)=k^2(1+3\chi_1+\chi_2), \\
3k_2^2&=&(\hat \xi_1-2\hat \xi_3)=k^2(1-2\chi_2), \\
3k_3^2&=&(\hat \xi_1-3\hat \xi_2+\hat \xi_3)=k^2(1-3\chi_1+\chi_2). 
\end{eqnarray}

\begin{figure}[htbp]
\includegraphics[scale=0.45]{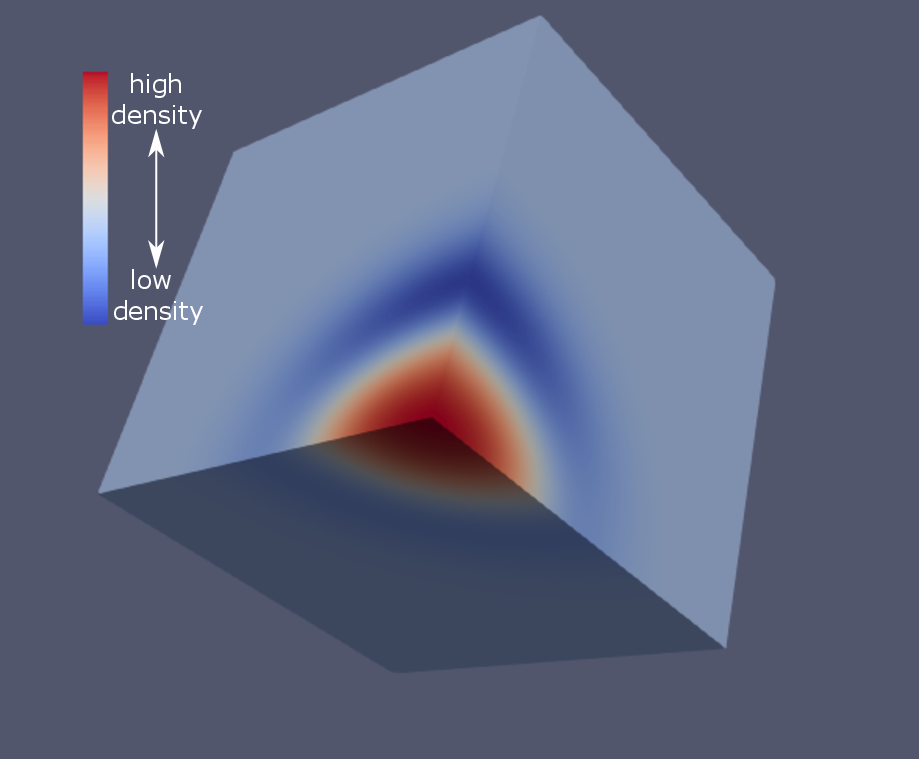}
\caption{The fluid comoving density 
at the initial time for $\mu=0.8$ and $\chi_2=0.1$. 
}
\label{fig:density}
\end{figure}

Let us summarize the physical parameters characterizing the initial data. 
First, we set the initial scale factor to 1. 
Taking $L$ as the unit of the length scale, we set 
$1/k=L/10$. 
The initial time slice is chosen so that it has 
a constant mean curvature $K_0$ 
by using the gauge degree of freedom. 
Then the initial Hubble parameter $H_0:=-K_0/3$ is chosen so that $1/H_0 =L/50=1/(5k)$, 
namely the scale of the inhomogeneity $1/k$ is 5 times larger than the initial Hubble length $1/H_0$. 
In this letter, we focus on the 
spheroidal profiles of the curvature perturbation, which are given by $\chi_1=|\chi_2|$. 
Then finally we have two free parameters $\mu$ and $\chi_2$.
The positive (negative) value of $\chi_{2}$ stands for 
oblateness (prolateness). 
In Fig.~\ref{fig:density}, we show the fluid comoving density 
at the initial time for $\mu=0.8$ and $\chi_2=0.1$. 

\section{Numerical schemes}
\label{sec:numesch}

For the simulation 
we use the 4th-order Runge-Kutta method with 
the BSSN~(Baumgarte-Shapiro-Shibata-Nakamura) formalism~\cite{Shibata:1995we,Baumgarte:1998te}
with the same gauge condition as in Ref.~\cite{Yoo:2013yea} 
and 
a central scheme with 
MUSCL~(Mono Upstream-centered Scheme for Conservation Laws)~\cite{2000JCoPh.160..241K,Shibata:2005jv} method 
for the fluid dynamics. 
Since we are interested in a cosmological setting, 
the boundary condition cannot be asymptotically flat. 
If spherical symmetry is imposed, 
we may use the asymptotic Friedmann-Lema\^itre-Robertson-Walker~(FLRW) 
condition~\cite{Shibata:1999zs} or 
just cut out the outer region causally connected to the outer boundary 
taking a sufficiently large initial region. 
However, the validity of the asymptotic FLRW condition is not clear in general, 
and the cutting-out procedure is not available due to the limited computational resources. 
Therefore we adopt the periodic boundary condition 
as is imposed in Refs.~\cite{Yoo:2013yea,Yoo:2018pda}. 
In this setting, we need to simultaneously resolve 
the scales of the gravitational collapse and cosmological expansion. 
In order to overcome this difficulty, 
for the spatial coordinates, we use  
the scale-up coordinates introduced in Ref.~\cite{Yoo:2018pda} with the parameter $\eta=15$, 
where 
the ratio between the Cartesian coordinate lengths of the unit coordinate interval at the boundary and origin
is $1+2\eta$. 
In the initial stage of the evolution, 
the typical time scale should be given by $1/|K|$ with $K$ being the trace of the extrinsic curvature 
at the point $x=y=z=L$. 
Thus we fix the time interval $\Delta t$ of the simulation by 
\begin{equation}
\Delta t=C\times \min\left\{\Delta x, 1/(10|K|)\right\}, 
\end{equation}
where we set the spatial grid interval $\Delta x=1/100$ and $C=1/20$.

\section{Results}
\label{sec:res}

\subsection{Spherical initial data}

\begin{figure}[htbp]
\includegraphics[scale=0.65]{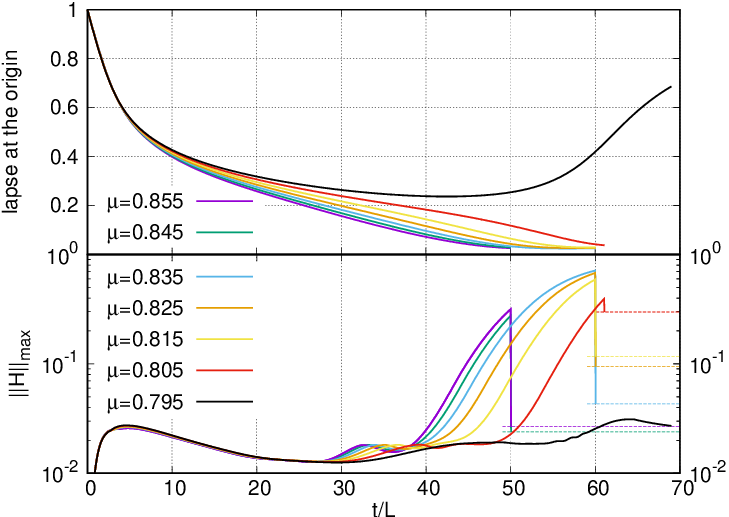}
\caption{The value of 
the lapse function at the origin~(upper) and 
the max norm of the Hamiltonian constraint~(lower) 
as functions of the time for each parameter set. 
}
\label{fig:alp_sph}
\end{figure}

\begin{figure}[htbp]
\includegraphics[scale=0.52]{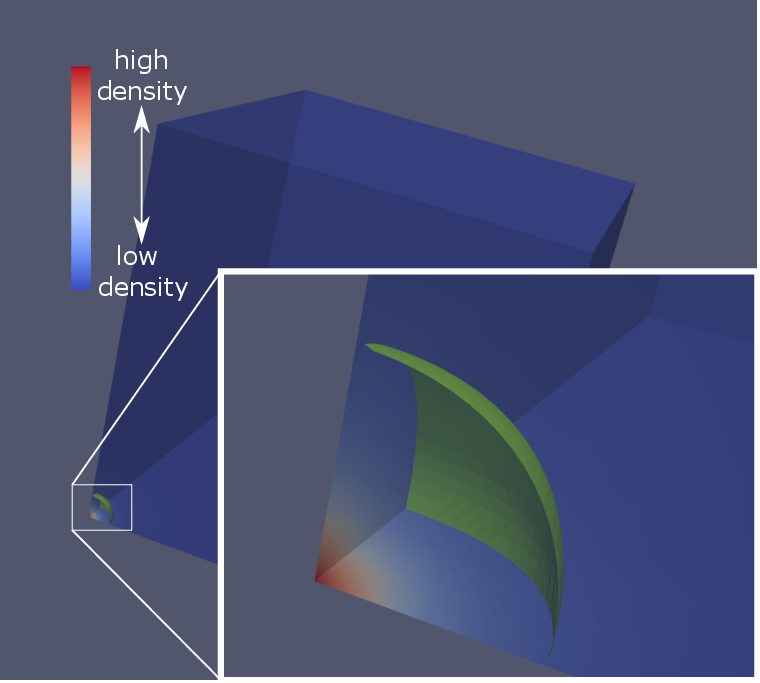}
\caption{Apparent horizon and the fluid comoving density 
at the time of the horizon formation for $\mu=0.805$. 
}
\label{fig:horizon}
\end{figure}

For the spherically symmetric cases, 
we find that the threshold value $\mu_{\rm th}$ is around 0.8. 
For $\mu\le0.795$, the collapse stops and bounces back. 
We can check this behavior from the time evolution of the value of the lapse function 
at the origin~(Fig.~\ref{fig:alp_sph}). 
On the other hand, for $\mu\ge0.805$, the fluid does not bounce back and 
finally we find an apparent horizon in the center~(Fig.~\ref{fig:horizon}). 
We also show the time evolution of 
the max norm of the Hamiltonian constraint violation $||H||_{\rm max}$
in Fig.~\ref{fig:alp_sph}. 
The function $||H||$ is normalized so that $||H||\leq1$ at each grid point. 
We take a
maximum over the whole computing region as $||H||_{{\rm max}}$
before the horizon formation, while we switch 
from the whole region to 
outside the horizon after the horizon formation. 
This switch gives a discontinuous reduction in $||H||_{{\rm max}}$ 
as seen in the lower panel of 
Fig.~\ref{fig:alp_sph} because $||H||$ takes a maximum in the very
central region before the horizon formation and the central region gets hidden behind 
the horizon after the horizon formation. 
If the value of $||H||_{\rm max}$ after the discontinuous reduction, which 
is denoted by the horizontal dashed line in the lower panel of 
Fig.~\ref{fig:alp_sph}, is well controlled, we can 
regard the computation outside the horizon acceptable.  
Fig.~\ref{fig:alp_sph} shows that 
even after the reduction, 
the constraint is significantly violated~($||H||_{\rm max}\sim 0.4$) 
around the horizon for the $\mu=0.805$ case. 
For $\mu \geq 0.845$,
however,  
we find that $||H||_{\rm max}$  
is well suppressed outside the apparent horizon at the time when we detect the horizon.
For $0.805\lesssim\mu\lesssim0.845$, we need more effort to resolve
the horizon formation. 
On the other hand, since $||H||_{\rm max}$ is always well controlled ($\lesssim 0.03$) for
the bouncing dynamics for $\mu=0.795$, 
we expect that the threshold value is given by $\mu_{\rm th}\simeq 0.8$. 

\begin{figure}[htbp]
\includegraphics[scale=0.7]{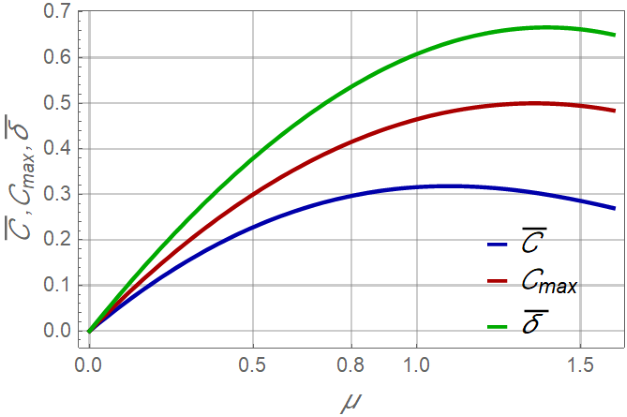}
\caption{The averaged compaction function $\bar {\mathcal C}$, 
the maximum compaction function ${\mathcal C}_{\rm max}$ and the averaged comoving density perturbation 
$\bar \delta$ in the overdense region at horizon entry as functions of $\mu$. 
}
\label{fig:deltaCC}
\end{figure}

Since the system is spherical 
if we ignore the effect of the boundary condition,  
we can check the resultant threshold value 
based on the compaction function $\mathcal C$ in the constant mean curvature slice~\cite{Shibata:1999zs}, 
which is directly related to the more conventional indicator $\bar \delta$,
the averaged density perturbation in the overdense region on the comoving slicing at horizon entry, through 
$\bar \delta=(4/3)\mathcal C$ if the radius for $\mathcal C$ 
is identified with that of $\bar \delta$~\cite{Harada:2015yda}. 
The threshold value $\sim 0.4$ of the maximum value $\mathcal C_{\rm max}$ 
is conventionally used. 
More recently, it has been reported that 
the volume average $\bar{\mathcal C}$ of $\mathcal C$ within the radius $r_{\rm m}$, 
at which $\mathcal C$ takes a maximum, 
gives a very stable threshold value of 0.3 
at a level of a few \% accuracy 
for a moderate shape of the inhomogeneity~\cite{Escriva:2019phb}. 
In Fig.~\ref{fig:deltaCC}, we show the values of 
$\bar {\mathcal C}$, ${\mathcal C}_{\rm max}$ and 
$\bar \delta$ 
as functions of $\mu$. 
For $\mu=0.8$, in our initial setting, 
the value of $\bar {\mathcal C}$ is given by 0.297 which is 
about only 1\% deviation from the reference value 0.3. 
Having this agreement, throughout this letter, 
we conclude that a PBH is formed if the bouncing-back behavior is not observed. 
For all the non-bouncing cases, even if the value of $||H||_{\rm max}$
becomes of the order of 1, 
we can eventually find an apparent horizon. 

\subsection{Non-spherical initial data}

\begin{figure}[htbp]
\includegraphics[scale=0.65]{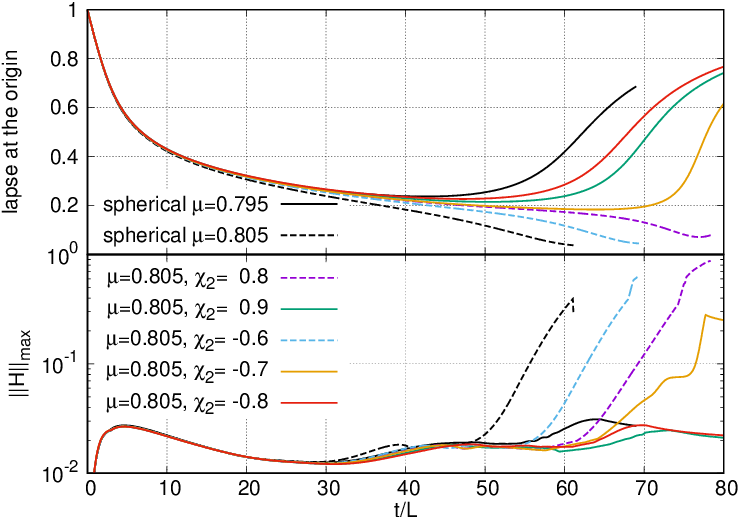}
\caption{The value of the lapse function at the origin~(upper) 
and the max norm of the Hamiltonian constraint~(lower) 
as functions of the time for each parameter set. 
The dashed lines are corresponding to the cases with horizon formation. 
}
\label{fig:nonsph_tdep}
\end{figure}

By numerical simulations with nonzero $\chi_{2}$, we find that the PBH formation becomes harder 
for larger ellipticity, which is consistent with the hoop conjecture~\cite{1972mwm..book.....K}. 
We look for the critical value of $\chi_2$ beyond or below which 
no horizon is formed,  for $\mu=0.805$. 
As a result, 
we find PBH formation for $-0.06\leq\chi_2\leq0.08$ with $\mu=0.805$, while we find a bouncing behavior for $\chi_2\leq -0.07$ or $\chi_2\geq 0.09$ (see Fig.~\ref{fig:nonsph_tdep}). 
Although we find a bouncing behavior for $\chi_2=-0.07$, since the value of $\chi_2$ is 
very close to the critical value, 
the Hamiltonian constraint is significantly violated near the center similarly to the collapsing cases. 
Unless $-0.08<\chi_2<0.09$, the constraint violation is well suppressed.

\begin{figure}[htbp]
\includegraphics[scale=0.9]{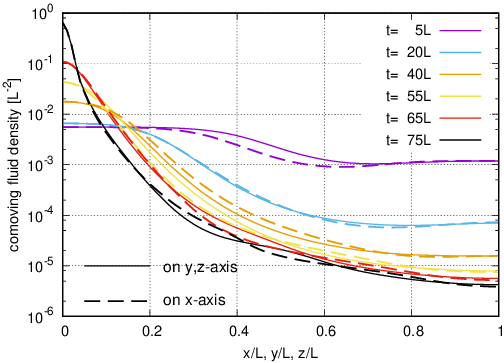}
\caption{Time evolution of 
the comoving fluid density
on each axis for $\chi_2=0.08$. 
}
\label{fig:nonsph_8}
\end{figure}

\begin{figure}[htbp]
\includegraphics[scale=0.9]{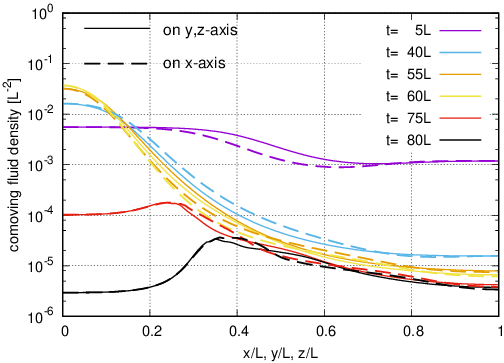}
\caption{Time evolution of the 
comoving fluid density 
on each axis for $\chi_2=0.09$. 
}
\label{fig:nonsph_9}
\end{figure}

\begin{figure}[htbp]
\includegraphics[scale=0.65]{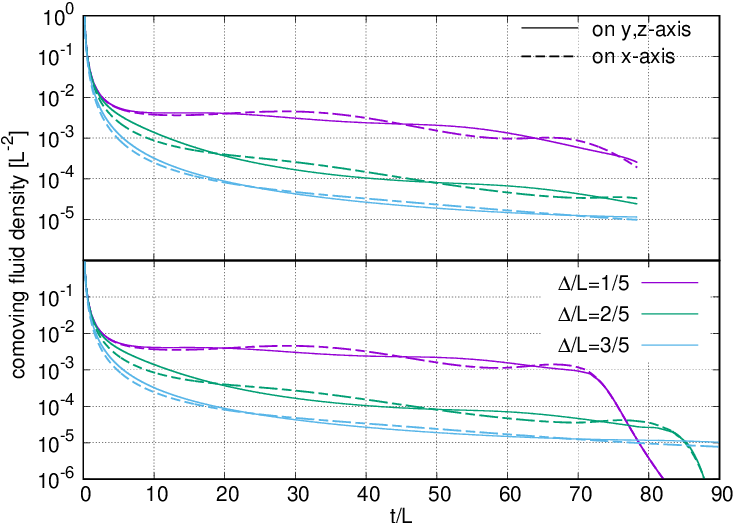}
\caption{Time evolution of the 
comoving fluid density at fixed spatial points on axes 
for $\chi_2=0.08$~(upper) and $\chi_2=0.09$~(lower), 
where $\Delta$ is the coordinate distance 
from the origin. 
}
\label{fig:osc_8}
\end{figure}

In Figs.~\ref{fig:nonsph_8} and \ref{fig:nonsph_9}, we show the time evolution of the comoving fluid density 
on each axis for $\chi_2=0.08$ and $0.09$ cases, respectively. 
For both cases, in  late times, 
the configuration is highly spherically symmetric near the center. 
We also find an oscillatory behavior between
prolateness and oblateness, which is expected by the result of the  
linear analysis of the nonspherical 
perturbations around the spherically symmetric critical 
solution reported in Ref.~\cite{Gundlach:1999cw}. 
The oscillation is more apparent in Fig.~\ref{fig:osc_8}, 
where the values of the comoving fluid density at specific spatial points 
are given as functions of the time. 
Therefore we conclude that the system is stable against the non-spherically symmetric perturbation of the current setting, 
whereas it slightly changes the threshold value of 
$\mu$ for the PBH formation.

\section{Summary and discussion}
\label{sec:sand}

We have performed numerical simulations for PBH formation 
for several values of $\chi_2$, 
which characterizes the initial spheroidal profile. 
It has been shown that the value of $\chi_2\sim 0.1$ gives only $\sim 1\%$ difference 
in the threshold value of the amplitude of the curvature perturbation $\mu$, 
and we typically have $|\chi_2|\ll 1$ for PBH formation in the radiation dominated universe. 
%
Thus we conclude that the effect of ellipticity on the threshold of PBH formation is 
highly limited and usually negligible 
in standard situations for PBH formation 
in the radiation-dominated universe.

As is expected from the general nature of the critical collapse,  
the final fate is sensitive to the parameters $\mu$ and $\chi_2$ around the critical values, 
so that the time evolution can be clearly classified with the existence of the bouncing behavior. 
Thus we can read off the threshold value and discuss the effect of the ellipticity 
although the resolution is not fine enough to resolve the horizon in our simulation. 
In order to analyze the finer structure of the solutions around the criticality, 
we need a finer resolution near the center. 
If the equation of state of the matter field is softer than the radiation fluid, 
the result would drastically change~(see Refs.~\cite{1982SvA....26....9P,Harada:2016mhb,Harada:2017fjm,Kokubu:2018fxy} for the pressureless matter). 
In order to analyze the spin generation of PBH, we 
have to consider the initial setting in which the tidal torque works during the collapse~\cite{DeLuca:2019buf}. 
These are beyond the scope of this letter and are left as future issues. 

\begin{acknowledgments}
This work was supported by JSPS KAKENHI Grant
Numbers JP19H01895(C.Y. and T.H), and JP19K03876 (T.H.), and in part  
by Waseda University Grant for Special Research Projects(Project number: 2019C-640).
\end{acknowledgments}


\begin{thebibliography}{10}

\bibitem{1967SvA....10..602Z}
Y.~B. {Zel'dovich} and I.~D. {Novikov},
\newblock Soviet Ast. {\bf 10}, 602 (1967).

\bibitem{Hawking:1971ei}
S.~Hawking,
\newblock Mon. Not. Roy. Astron. Soc. {\bf 152}, 75 (1971).

\bibitem{Carr:2020gox}
B.~Carr, K.~Kohri, Y.~Sendouda, and J.~Yokoyama,
\newblock (2020), arXiv:2002.12778.

\bibitem{Abbott:2016blz}
Virgo, LIGO Scientific, B.~P. Abbott {\it et~al.},
\newblock Phys. Rev. Lett. {\bf 116}, 061102 (2016), arXiv:1602.03837.

\bibitem{Sasaki:2016jop}
M.~Sasaki, T.~Suyama, T.~Tanaka, and S.~Yokoyama,
\newblock Phys. Rev. Lett. {\bf 117}, 061101 (2016), arXiv:1603.08338.

\bibitem{Carr:1975qj}
B.~J. Carr,
\newblock Astrophys. J. {\bf 201}, 1 (1975).

\bibitem{Harada:2013epa}
T.~Harada, C.-M. Yoo, and K.~Kohri,
\newblock Phys. Rev. {\bf D88}, 084051 (2013), arXiv:1309.4201,
\newblock [Erratum: Phys. Rev.D89,no.2,029903(2014)].

\bibitem{1978SvA....22..129N}
D.~K. {Nadezhin}, I.~D. {Novikov}, and A.~G. {Polnarev},
\newblock Soviet Ast. {\bf 22}, 129 (1978).

\bibitem{1980SvA....24..147N}
I.~D. {Novikov} and A.~G. {Polnarev},
\newblock Soviet Ast. {\bf 24}, 147 (1980).

\bibitem{Shibata:1999zs}
M.~Shibata and M.~Sasaki,
\newblock Phys. Rev. {\bf D60}, 084002 (1999), arXiv:gr-qc/9905064.

\bibitem{Niemeyer:1999ak}
J.~C. Niemeyer and K.~Jedamzik,
\newblock Phys. Rev. {\bf D59}, 124013 (1999), arXiv:astro-ph/9901292.

\bibitem{Musco:2004ak}
I.~Musco, J.~C. Miller, and L.~Rezzolla,
\newblock Class. Quant. Grav. {\bf 22}, 1405 (2005), arXiv:gr-qc/0412063.

\bibitem{1981ApJ...250..432B}
J.~D. {Barrow} and J.~{Silk},
\newblock \apj {\bf 250}, 432 (1981).

\bibitem{Kuhnel:2016exn}
F.~K\"uhnel and M.~Sandstad,
\newblock Phys. Rev. {\bf D94}, 063514 (2016), arXiv:1602.04815.

\bibitem{Chiba:2017rvs}
T.~Chiba and S.~Yokoyama,
\newblock PTEP {\bf 2017}, 083E01 (2017), arXiv:1704.06573.

\bibitem{Harada:2017fjm}
T.~Harada, C.-M. Yoo, K.~Kohri, and K.-I. Nakao,
\newblock Phys. Rev. {\bf D96}, 083517 (2017), arXiv:1707.03595,
\newblock [Erratum: Phys. Rev.D99,no.6,069904(2019)].

\bibitem{DeLuca:2019buf}
V.~De~Luca, V.~Desjacques, G.~Franciolini, A.~Malhotra, and A.~Riotto,
\newblock JCAP {\bf 1905}, 018 (2019), arXiv:1903.01179.

\bibitem{Mirbabayi:2019uph}
M.~Mirbabayi, A.~Gruzinov, and J.~Nore\~na,
\newblock (2019), arXiv:1901.05963.

\bibitem{Fernandez:2019kyb}
N.~Fernandez and S.~Profumo,
\newblock JCAP {\bf 1908}, 022 (2019), arXiv:1905.13019.

\bibitem{He:2019cdb}
M.~He and T.~Suyama,
\newblock Phys. Rev. {\bf D100}, 063520 (2019), arXiv:1906.10987.


\bibitem{Harada:2015yda}
T.~Harada, C.-M. Yoo, T.~Nakama, and Y.~Koga,
\newblock Phys. Rev. {\bf D91}, 084057 (2015), arXiv:1503.03934.

\bibitem{Lyth:2004gb}
D.~H. Lyth, K.~A. Malik, and M.~Sasaki,
\newblock JCAP {\bf 0505}, 004 (2005), arXiv:astro-ph/0411220.

\bibitem{1986ApJ...304...15B}
J.~M. {Bardeen}, J.~R. {Bond}, N.~{Kaiser}, and A.~S. {Szalay},
\newblock \apj {\bf 304}, 15 (1986).

\bibitem{Yoo:2018kvb}
C.-M. Yoo, T.~Harada, J.~Garriga, and K.~Kohri,
\newblock (2018), arXiv:1805.03946.

\bibitem{Yoo:2013yea}
C.-M. Yoo, H.~Okawa, and K.-i. Nakao,
\newblock Phys. Rev. Lett. {\bf 111}, 161102 (2013), arXiv:1306.1389.

\bibitem{Yoo:2018pda}
C.-M. Yoo, T.~Ikeda, and H.~Okawa,
\newblock Class. Quant. Grav. {\bf 36}, 075004 (2019), arXiv:1811.00762.

\bibitem{Shibata:1995we}
M.~Shibata and T.~Nakamura,
\newblock Phys.Rev. {\bf D52}, 5428 (1995).

\bibitem{Baumgarte:1998te}
T.~W. Baumgarte and S.~L. Shapiro,
\newblock Phys.Rev. {\bf D59}, 024007 (1999), arXiv:gr-qc/9810065.

\bibitem{2000JCoPh.160..241K}
A.~{Kurganov} and E.~{Tadmor},
\newblock Journal of Computational Physics {\bf 160}, 241 (2000).

\bibitem{Shibata:2005jv}
M.~Shibata and J.~A. Font,
\newblock Phys. Rev. {\bf D72}, 047501 (2005), arXiv:gr-qc/0507099.

\bibitem{Escriva:2019phb}
A.~Escriv\`a, C.~Germani, and R.~K. Sheth,
\newblock Phys. Rev. {\bf D101}, 044022 (2020), arXiv:1907.13311.

\bibitem{1972mwm..book.....K}
K.~S. Thorne,
\newblock {\it Magic Without Magic} ,
\newblock (1972), John Archibald Wheeler, John R. Klauder(eds.), Freeman, San
  Fransisco.

\bibitem{Gundlach:1999cw}
C.~Gundlach,
\newblock Phys. Rev. {\bf D65}, 084021 (2002), arXiv:gr-qc/9906124.

\bibitem{1982SvA....26....9P}
A.~G. {Polnarev} and M.~Y. {Khlopov},
\newblock Soviet Ast. {\bf 26}, 9 (1982).

\bibitem{Harada:2016mhb}
T.~Harada, C.-M. Yoo, K.~Kohri, K.-i. Nakao, and S.~Jhingan,
\newblock Astrophys. J. {\bf 833}, 61 (2016), arXiv:1609.01588.

\bibitem{Kokubu:2018fxy}
T.~Kokubu, K.~Kyutoku, K.~Kohri, and T.~Harada,
\newblock Phys. Rev. {\bf D98}, 123024 (2018), arXiv:1810.03490.

\end{thebibliography}
\end{document}